\documentclass[11pt]{article}

\usepackage{graphicx}
\usepackage{floatflt}

\newcommand {\re}{{\rm e}}\newcommand {\ri}{{\rm i}}
\newcommand {\rK}{{\rm K}}

\newcommand {\rBz}{{\rm B^0}} \newcommand {\rBqz}{{\overline{\rm B}}{}^0}
\newcommand {\rBs}{{\rm B_s}} \newcommand {\rBqs}{\rm{\overline B}{}_s}
\newcommand {\rDz}{{\rm D^0}} \newcommand {\rDqz}{{\overline{\rm D}}{}^0}
\newcommand {\rKz}{{\rm K^0}} \newcommand {\rKqz}{{\overline{\rm K}}{}^0}

\newcommand {\Begeq}{\begin{equation}}
\newcommand {\Endeq}{\end{equation}}
\newcommand {\bEa}{\begin{eqnarray}}
\newcommand {\eEa}{\end{eqnarray}}

\begin{document}

\begin{center}
{\bf{\large{
Translation of Time-Reversal Violation in the Neutral K-Meson System into
a Table-Top Mechanical System
}}}\\[8mm]
%%%
Andreas Reiser$^1$, Klaus R. Schubert$^2$, and J\"urgen Stiewe$^1$\\
%%%
$^1$ Kirchhoff-Institut f\"ur Physik, Universit\"at Heidelberg,
Heidelberg, Germany\\
%%%
$^2$ Institut f\"ur Kern- und Teilchenphysik, Technische Universit\"at
Dresden,\\ Dresden, Germany\\

%%% 
\vspace{3mm}
E-mail: Andreas.Reiser@kip.uni-heidelberg.de\\[3mm]

\end{center}

\begin{abstract}

\noindent
Weak interactions break time-reversal (T) symmetry in the
two-state system of neutral K mesons. We present and discuss a
two-state mechanical system, a Foucault-type pendulum on a
rotating table, for a full representation of 
${\rKz}{\rKqz}$ transitions
by the pendulum motions including T violation. The pendulum
moves with two different oscillation frequencies and two different
magnetic dampings. Its equation of motion is identical with
the differential equation for the real part of the CPT-symmetric K-meson
wave function. The pendulum is able to represent microscopic CP and
T violation with CPT symmetry owing to the macroscopic Coriolis force
which breaks the symmetry under reversal-of-motion. Video clips of
the pendulum motions are shown as supplementary material.

\end{abstract}

\vspace{3mm}

\noindent
{\bf I. Introduction}\\

\noindent
Symmetry considerations belong to the most fundamental and powerful tools in many fields of physics. Especially in particle physics, symmetries both under continuous and under discrete transformations play a very important role. Strong and electromagnetic interactions are symmetric under the three discrete transformations parity 
P $(\vec{r} \leftrightarrow - \vec{r})$, charge conjugation C ( particle $\leftrightarrow$ antiparticle) and time reversal T (t $\leftrightarrow$~-t and initial  $\leftrightarrow$ final states). As observed in 1957  
\cite{Wu, Lederman},
 weak interactions break the discrete symmetries P and C maximally, but they are symmetric under the combination CP to a high level of accuracy.  In 1964, a violation of CP symmetry at the level of 
$10^{-3}$ has been observed in transitions between ${\rKz}$
and ${\rKqz}$ mesons \cite{Christenson}. 
Today we understand that these transitions and the small symmetry breaking therein are also produced by weak interactions. The Standard Theory of weak interactions is symmetric under the combination CPT. In this theory, CP violation goes along with T violation and is allowed to show up in any weak process involving three families of quarks \cite{Kobayashi}. T violation 
and CPT symmetry in 
${\rKz}{\rKqz}$ transitions  has been confirmed by experiments
\cite{1970-Schubert, 1998-CPLEAR, 2006-FidecaroGerber}, 
and no CPT-symmetry breaking process has been observed so far.\\

\noindent
Using the present precise values of the 
${\rKz}{\rKqz}$-transition parameters \cite{2010-PDG}, figure 1 shows the time-dependent probabilities for (a) an initial 
${\rKz}$ remaining a ${\rKz}$ and appearing as a 
${\rKqz}$, and for (b) an initial ${\rKqz}$  remaining a 
${\rKqz}$  and appearing as a ${\rKz}$. For better visibility, the size of the observed T violation has been increased by a factor of 10. The equality of the two pobabilities 
$P[{\rKz}(0) \to {\rKz}(t)]$ and $P[{\rKqz}(0) \to {\rKqz}(t)]$ 
means CPT symmetry, and the
inequality 
$P[{\rKqz}(0) \to {\rKz}(t)] > P[{\rKz}(0) \to {\rKqz}(t)]$ 
breaks CP as well as T symmetry. The two sums 
$P[{\rKqz}(0) \to {\rKz}(t)$ or ${\rKqz}(t)]$ 
and $P[{\rKz}(0) \to {\rKz}(t)$ or ${\rKqz}(t)]$ in (c)  
show that initial ${\rKqz}$ ``live longer'' than initial 
${\rKz}$ mesons, i.e. have produced a smaller number of decays in the same interval of time.\\

%%%
\begin{figure}[h]\begin{center}
\includegraphics[width=8cm]{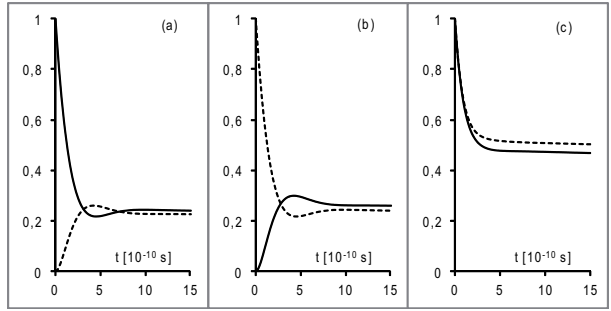}\end{center}
\caption[ ]{In (a) for a ${\rKz}$ at t = 0 and in (b) for a 
${\rKqz}$ at t = 0, the solid curves show the 
${\rKz}$ intensities and the dashed curves the 
${\rKqz}$ intensities. In (c), the solid curve is the sum of the two curves in (a), and the dashed curve is the sum of the two curves in (b).}
\end{figure}
%%%

\noindent
Several authors have proposed and constructed mechanical and electrical 
analoga of the 
${\rKz}{\rKqz}$ system starting, 
to our knowledge, with Winstein in 1987. He constructed and presented a pair of coupled pendula swinging in one dimension. In the writeup of his presentation  \cite{1988-Winstein}, Winstein 
proved that his analogon is principally not able to show T violation because of the time-reversal symmetry of the equations of motion for the pendulum pair. In 1996, Rosner \cite{1996-Rosner} 
proposed that the Coriolis force can be used for demonstrating T violation in 
${\rKz}{\rKqz}$
transitions with one pendulum swinging in two dimensions on a turntable. Kostelecki and Roberts \cite{2001-KosteleckyRoberts}
discussed this further and also proposed to use the T violating properties of a gyrator in a setup with two coupled electrical circuits. The latter proposal has recently been discussed again by Caruso et al. \cite{2011-Caruso}. 
We are not aware that one of these T violating proposals has been built so far and we therefore present in this article
the construction of a table-top Foucault pendulum.\\ 

\noindent
Our setup follows the suggestion of Rosner and Slezak 
\cite{2001-RosnerSlezak} with anisotropic oscillations and dampings of the pendulum. In chapter II we describe the construction, in 
chapter III we present the equation of motion and its solutions. In chapter IV we show that the real parts of the Schr\"odinger-equation solutions for the 
${\rKz}{\rKqz}$ system obey a second-order 
differential equation which is formally identical with that for the coordinates of the pendulum. Therefore, as shown previously for the pair of coupled pendula 
\cite{2012-SchubertStiewe}
and by Caruso et al. \cite{2011-Caruso}, the table-top Foucault pendulum is not only demonstration and illustration of the 
${\rKz}{\rKqz}$ system, 
but also isomorphic mapping, ``translation'' or equivalence in a stricter
sense. The effect of the macroscopically T-violating Coriolis force is able to represent the microscopic T violation in 
${\rKz}{\rKqz}$ transitions. In chapter V, after discussing some control measurements, we present the motions of the pendulum and add some extensions. Six video clips of the motions are added to the article as supplementary material
\footnote{http://www.kip.uni-heidelberg.de/user/reiser/video}.
\newpage

%\vfill
%%%
\begin{figure}[h]\begin{center}
\includegraphics[width=5.8cm]{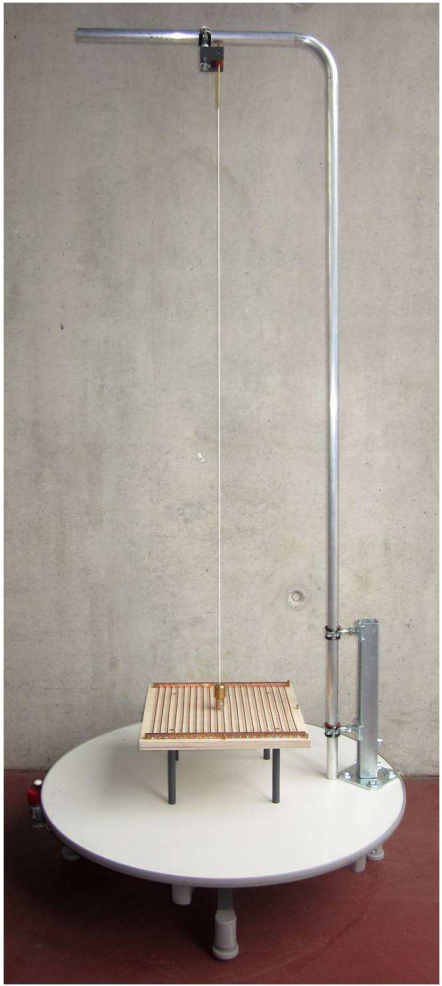}\end{center}
\caption[ ]{Photograph of the pendulum setup}
\end{figure}
%%%
%\vfill

\vspace{5mm} 

\noindent
{\bf II.  Apparatus - A Foucault Pendulum on a Turntable}\\

\noindent The Foucault pendulum for representing T violation is constructed as a portable system; figure 2 shows a photograph 
and figure 3 a drawing of its main components. 
The rotating frame of reference is a wooden circular plate with a diameter of 1 m. Since there is no preferred direction in the rotating plane, we choose an arbitrary direction as x-axis representing the 
${\rKz}$ state. The perpendicular direction y represents the 
${\rKqz}$ state. The axial vector $\Omega\,\vec{e}_z$
describes the rotation, and $\vec{e}_x, \vec{e}_y, \vec{e}_z$ form a 
right-handed orthogonal coordinate system. 
The turntable is mounted  on a steel construction which has four adjustable feet at the lower part and a ball-borne axis carrying four upper arms. The wooden plate is fixed on these arms. We use an electric gear dc motor to drive the turntable. The motor is mounted at one of the lower arms. Its axis carries a rubber wheel. The complete drive unit is pressed to the circumference of the wooden table using a spring. This ensures a good frictional contact even if the wooden plate is not perfectly  circular. The accessible rotation speed of the turntable ranges between 0 and $0.1$ rpm and can be adjusted by setting the appropriate drive voltage for the motor. Fixed at an aluminium stand
mounted onto the turntable, a non-twisting nylon thread 
of length $\ell$ = 1.35~m carries the pendulum bob. 
The upper part of the thread can only swing back and forth in (x+y) direction enforced by a hinge and a tube of length $\delta \ell$ = 10 cm fixed to the hinge. The lower part of the thread can swing freely in any direction. For motions in the (x-y) direction we have  
$\Omega_- = \sqrt{g/(\ell-\delta \ell)}$ , 
in the (x+y) direction $\Omega_+ = \sqrt{g/\ell}$.  The pendulum bob is made
of a cylindrical brass weight, $30 \, \mathrm{mm}$ long and $30 \, \mathrm{mm}$ in diameter. The pendulum thread is fixed using a centric bore of  $1 \, \mathrm{mm}$ diameter in the brass weight. 
One ring-shaped permanent dipolar magnet with an outer diameter of  $20 \, \mathrm{mm}$, inner diameter of $10 \, \mathrm{mm}$ and height $10 \, \mathrm{mm}$ is glued to the brass weight using an epoxy resin. 
The material is an alloy (N45) made of NdFeB, with a maximum 
field of approximately $1.37\, \mathrm{T}$ at the surface. The lower end of the brass weight is machined to fit tightly into the bore of the magnet. A second N45 magnet, disc-shaped with an diameter of $20 \, \mathrm{mm}$ and height $10 \, \mathrm{mm}$ forms the closure of the pendulum bob and enlarges the magnetic field. The dipole moment of each magnet is oriented along the cylinder axis.\\

\begin{figure}[h]\begin{center}
\includegraphics[width=14cm]{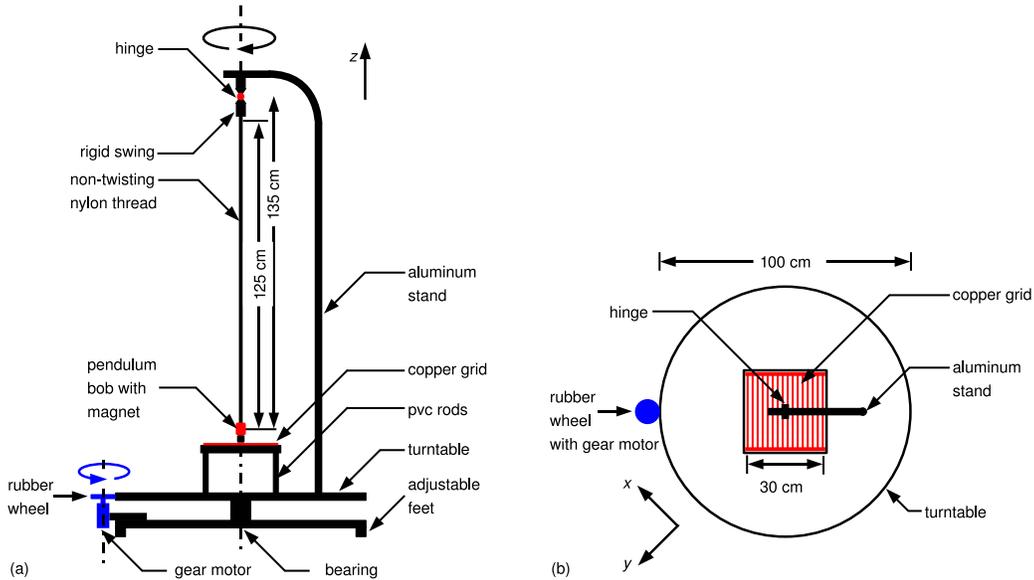}\end{center}
\caption[ ]{Side (a) and top (b) view of the complete pendulum setup}
\end{figure}

\noindent
Supported by four $15 \, \mathrm{cm}$ long pvc rods, a square wooden plate  of $33 \, \mathrm{cm}$ edge length is used to carry the damping device. This small table is screwed at the turntable 
centrically. The damping device consists of a copper-rod grid in which eddy currents are induced by the magnetic bob. The copper rods are $30 \, \mathrm{cm}$ long and have a diameter of $3 \, \mathrm{mm}$. The spacing between two copper rods is $1.5\, \mathrm{cm}$. The copper grid is closed at both ends via soft-soldering the copper rods together into a massive copper bar $31 \, \mathrm{cm}$ long and $10 \times 10 \, \mathrm{mm}$ in cross section. This low-ohmic electrical shortcut bar is necessary for an effective eddy current brake. The length of the pvc rods was chosen in order to avoid magnetic   
disturbance of the pendulum bob originating 
from the steel structure of the turntable. All materials close to the magnetic pendulum bob are non-magnetic in common sense (wood, copper, brass, pvc). The copper grid lies in the $xy$ plane   
and is aligned paralled to the $(x-y)$  direction. The magnetic pendulum bob is swinging above this grid with a minimum distance of $5 \, \mathrm{mm}$. The damping in $(x-y)$  direction is small whereas it is large in the $(x+y)$  direction. To explain the damping mechanism we consider the magnetic flux caused by the magnetic pendulum bob.
When moving the magnet along the rods the flux change in the shortcut loops of the grid is practically vanishing and no damping force  is present. A movement perpendicular to the copper rods leads to the maximum flux change within the loops causing a  damping force owing to Lenz' rule. With this setup, the damping force is approximately proportional to the bob velocity.\\

\noindent
{\bf III. Equation of Motion and its Solutions}\\

\noindent
The damped harmonic oscillator with a period $T=2\pi/\omega$ has
the equation of motion
\Begeq
{\ddot x} = - \omega^2 x -\Gamma {\dot x}~,\label{Eq:III.1}
\Endeq
where $1/\Gamma$ is the time in which its energy decreases by 1/e  
and $2/\Gamma$ the same for the amplitude. The solution of the equation is
\Begeq
x(t) = A \re^{-\Gamma t/2} \cos(\omega_0 t -B) \label{Eq:III.2}
\Endeq
with two free parameters $A$ and $B$, and $\omega_0=\sqrt{\omega^2
-\Gamma^2/4}$. With $\Omega=0$, 
two coordinates of the pendulum bob, $x-y$ and $x+y$, obey 
equation~(\ref{Eq:III.1}) if we neglect the z coordinate and restrict 
the motions 
to small deflection angles:
\bEa
{\ddot x-\ddot y} &=& - \omega_-^2 (x-y) -\Gamma_- (\dot x-\dot y)~,
\nonumber\\
{\ddot x+\ddot y} &=& - \omega_+^2 (x+y) -\Gamma_+ (\dot x+\dot y)~.
\eEa
The two equations are uncoupled. If we rotate the turntable, we have
to add the Coriois term $+ 2\Omega (\dot x +\dot y)$ to the first
equation and $- 2\Omega (\dot x -\dot y)$ to the second one, and we obtain 
a pair of coupled equations. The signs
of the Coriolis terms follow from the righthandedness of the system
$x-y,x+y,\Omega$. In $x$ and $y$, the coupled
equation of motion is
\Begeq
\left(\begin{array}{c}\ddot x\\ \ddot y\end{array}\right)=
-\left(\begin{array}{cc}\omega_0^2&\omega_0\cdot\Delta\omega\\
\omega_0\cdot\Delta\omega & \omega_0^2\end{array}\right)
\left(\begin{array}{c} x\\ y\end{array}\right)
-\left(\begin{array}{cc}\Gamma&\Delta\Gamma/2-2\Omega\\
\Delta\Gamma/2+2\Omega&\Gamma\end{array}\right)
\left(\begin{array}{c}\dot x\\ \dot y\end{array}\right)~, \label{Eq:III.3}
\Endeq
where we have introduced the abbreviations
\Begeq
\omega_0=(\omega_+ +\omega_-)/2~,~~\Delta\omega=\omega_+ -\omega_-~,~~
\Gamma = (\Gamma_+ +\Gamma_-)/2~,~~\Delta\Gamma=\Gamma_+ -\Gamma_-
\Endeq
and have used the approximation $|\Delta\omega|\ll\omega_0$.
Without further approximation, equation~(\ref{Eq:III.3}) has no
solutions of the form in equation~(\ref{Eq:III.2}). With the
approximation $|\Omega|\ll |\Delta\Gamma|$, however, the two eigensolutions
are of this form:
\Begeq
\left(\begin{array}{c}x\\ y\end{array}\right)_S = A
\left(\begin{array}{c}p\\ q\end{array}\right)
\re^{-\Gamma_S t/2} \cos(\omega_S t -B)~,~~
\left(\begin{array}{c}x\\ y\end{array}\right)_L = C
\left(\begin{array}{c}p\\ -q\end{array}\right)
\re^{-\Gamma_L t/2} \cos(\omega_L t -D)~,\label{Eq:III.5}
\Endeq
with 
\Begeq
\omega_S=\omega_+~,~~ \omega_L =\omega_-~,~~\Gamma_S = \Gamma_+~,~~
\Gamma_L=\Gamma_-~,~~p/q =1-4\Omega/\Delta\Gamma~,
\Endeq 
and $p^2+q^2=1$.
In our apparatus, we have chosen $\Gamma_+>\Gamma_-$, i.~e. we have
$\Delta\Gamma>0$ and the index $S$ and $L$ denotes, as in the Kaon system,
the short- and long-living eigenstate, respectively. For further equivalence with the Kaon
system, our demonstration has to use $\Delta\omega<0$ since the mass
difference between $\rKz_S$ and $\rKz_L$ is negative, and we have to turn
the table clockwise ($\Omega<0$) since $p/q=1+2 {\rm Re}(\epsilon_\rK)>1$.\\

\begin{figure}[h]\begin{center}
\includegraphics[width=8.5cm]{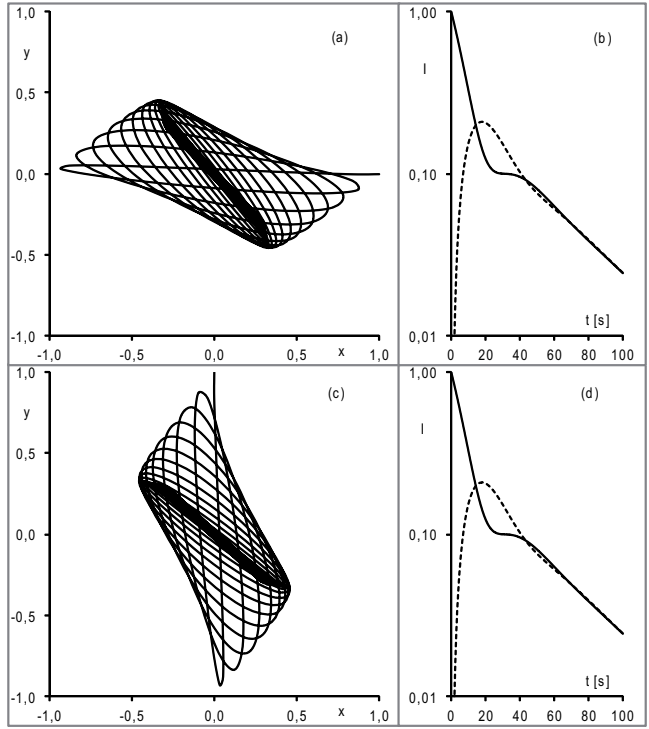}\end{center}
\caption[ ]{(a) Pendulum orbit with $\Omega = 0$ when starting the motion at $(x{,}y) = (1{,}0)$ for the
first 100 s. (b) The solid curve shows the intensity $< 2x^2 >$,
the dashed curve $< 2y^2 >$.
(c) Orbit with $\Omega = 0$ when starting at $(x{,}y)= (0{,}1)$. (d) Intensities 
$< 2y^2 >$ (solid curve)
and $< 2x^2 >$ (dashed curve) for the second motion; both curves are identical with those in (b). The parameters for the calculation are taken from the demonstration setup as given in chapter V. The asymptotic motions in (a) and (c) follow a straight line along the $(x-y)$ direction.}
\end{figure}

\noindent
The most general solution of equation~(\ref{Eq:III.3}), the sum of the
two eigenfunctions in equation~(\ref{Eq:III.5}), has four free parameters
since the equation of motion is of second order. The four parameters are
fixed by the initial conditions $x(0)$, $y(0)$, $\dot x(0)$ and 
$\dot y(0)$. The Schr\"odinger equation for the Kaon system is only of
first order, see chapter IV. Its most general solution contains only two 
free parameters to be fixed by $\psi_1(0)$ and $\psi_2(0)$. In the 
translation of the Kaon system into the pendulum, this means that we can
only fix $A$ and $C$ by $x(0)$ and $y(0)$, and the non-adjustable decay
rates $\dot x(0)$ and $\dot y(0)$ lead to $B=D=0$.\\

\noindent
With $\Omega=0$, figure 4 shows the orbits $(x,y)$ and intensities 
for two special solutions, with 
$(x,y)=(1,0)$ and $(x,y)=(0,1)$ at $t=0$; 
both are presented here with $B=D=0$. Since it is hard to implement 
$(\dot x,\dot y)=(-\Gamma/2,0)$ 
or $(0,-\Gamma/2)$ at $t$ = 0 in the demonstration of the pendulum, we have checked
that the $B$ and $D$ values originating from $(\dot x,\dot y)=(0,0)$
lead to orbits which show no noticeable difference from those with
$B=D=0$. In the first 20 seconds the orbit in figure 4(a), i.e. the translation of the evolution
of a pure $\rKz$ at $t=0$, shows a counterclockwise elliptical motion. Hereby the ellipse axes rotate clockwise. In figure 4(c), showing the evolution of a pure $\rKqz$, the two rotating senses are inverted.
This difference is not
seen in the intensities $\langle 2 x^2\rangle$ and $\langle 2 y^2\rangle$,
the intensities are equal. 
Because of the strong damping of the $(x+y)$ component of the 
motion, both orbits end in the less-damped linear motion in the $(x-y)$
direction. In this asymptotic motion, there are equal amplitudes in $x$ 
and $y$, and the remaining motion energy is the same for both orbits.
There is a complete symmetry between $x$ and $y$.\\

\begin{figure}[h!]\begin{center}
\includegraphics[width=8.5cm]{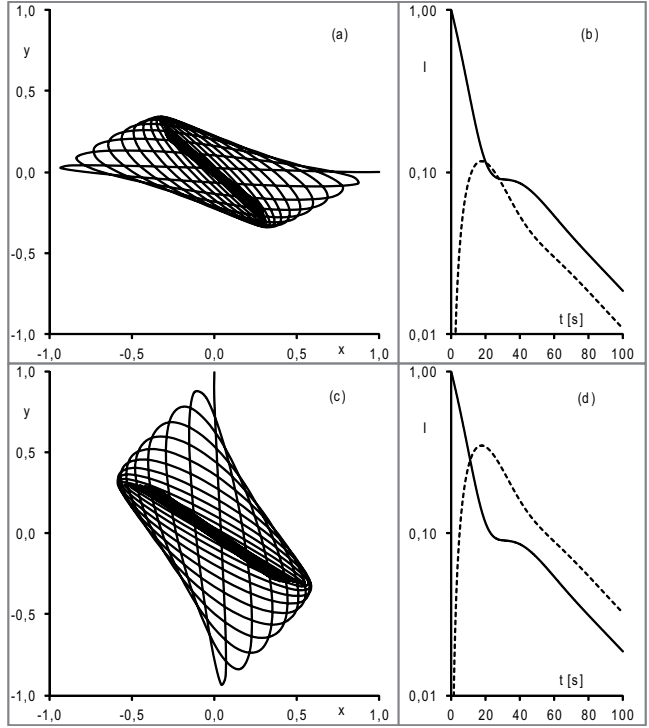}\end{center}
\caption[ ]{ (a) Pendulum orbit with $\Omega = -0.01/s$ when starting 
the motion at $(x{,}y) = (1{,}0)$ for the
first 100 s. (b) The solid curve shows the intensity $< 2x^2 >$, 
the dashed curve $< 2y^2 >$.
(c) Orbit with $\Omega = -0.01/s$ when starting at $(x{,}y)= (0{,}1)$. 
(d) Intensities $< 2y^2 >$ (solid curve)
and $< 2x^2 >$ (dashed curve) for the second motion. The asymptotic motions in (a) and (c) follow a straight line inclined by $7.6$ degrees from the $(x-y)$ direction.}
\end{figure}

\noindent
Figure 5 shows the orbits and intensities
for a clockwise table rotation with
$\Omega=-0.01$/s. The transition features
$x\to y$ and $y\to x$ are very much the same as for $\Omega=0$. The small 
observed difference is created by the T-violating Coriolis force. 
After damping
the S component of the motion, the asymptotic L motion has two pronounced 
features in breaking the symmetry 
between $x$ and $y$: First, the $x$ component is larger than the $y$
component for both initial conditions, and second, the remaining
motion energy at any time is larger when starting with $(x,y)=(0,1)$ than 
that for starting with $(x,y)=(1,0)$. Both features originate 
from the time-reversal violation $x^2(t)/y^2(0)>y^2(t)/x^2(0)$ created
by the Coriolis force.\\

\noindent
{\bf IV. Translating the Schr\"odinger Equation of the $\rKz\rKqz$ System}\\

\noindent
The time evolution of the wave function  
$\Psi=\psi_1 \rKz +\psi_2 \rKqz$ for the $\rKz\rKqz$ system
with CPT symmetry  
is given by the Schr\"odinger equation 
\Begeq 
\ri\frac{\partial}{\partial t }\left(\begin{array}{c}\psi_1\\
\psi_2\end{array}\right)=
\left[\left(\begin{array}{cc}m&m_{12}\\m_{12}^*&m\end{array}\right)
-\frac{\ri}{2}\left(\begin{array}{cc}\Gamma&\Gamma_{12}\\
\Gamma_{12}^*&\Gamma\end{array}\right)\right]
\left(\begin{array}{c}\psi_1\\\psi_2\end{array}\right) \label{Eq:A.1} 
\Endeq
with five observable real parameters, $m$, $\Gamma$, $|m_{12}|$,
${\rm Re}~(\Gamma_{12}/m_{12})$ and  ${\rm Im}~(\Gamma_{12}/m_{12})$.
T is violated iff ${\rm Im}~(\Gamma_{12}/m_{12}) \ne 0$.
The two fundamental solutions are
\Begeq
\left(\begin{array}{c}\psi_1\\ \psi_2\end{array}\right)_S = 
\left(\begin{array}{c}p\\ q\end{array}\right)
\cdot\re^{-\ri m_S t}~\re^{-\Gamma_St/2}~,~~
\left(\begin{array}{c}\psi_1\\ \psi_2\end{array}\right)_L = 
\left(\begin{array}{c}p\\ -q\end{array}\right)
\cdot\re^{-\ri m_Lt}~\re^{-\Gamma_Lt/2}~, \label{Eq:A.2}                        
\Endeq
where the five observables $m_S$, $\Gamma_S$, $m_L$, $\Gamma_L$
and $|p/q|$ follow unambiguously from the five parameters of the
Schr\"odinger equation \cite{BrancoLavouraSilva}. The values of $p$ and
$q$ can be chosen to be real, leading to
\Begeq
\left(\begin{array}{c}\Re_1\\ \Re_2\end{array}\right)_S = 
\left(\begin{array}{c}p\\ q\end{array}\right)
\cdot\re^{-\Gamma_St/2}~\cos(m_St)~,~~
\left(\begin{array}{c}\Re_1\\ \Re_2\end{array}\right)_L = 
\left(\begin{array}{c}p\\ -q\end{array}\right)
\cdot\re^{-\Gamma_Lt/2}~\cos(m_Lt)~.\label{Eq:A.3}                        
\Endeq
The real parts $\Re_1$ and $\Re_2$ of $\psi_1$ and $\psi_2$,
respecively, obey the second-order linear differential equation
\cite{2012-SchubertStiewe}
\bEa
\left(\begin{array}{c}\ddot\Re_1\\ \ddot\Re_2\end{array}\right)=-\left(
\begin{array}{cc}\frac{m_S^2+m_L^2}{2}+\frac{\Gamma_S^2+\Gamma_L^2}{8}&
\frac{p}{q}\cdot\left(\frac{m_S^2-m_L^2}{2}+\frac{\Gamma_S^2-\Gamma_L^2}{8}
\right)\\\frac{q}{p}\cdot
\left(\frac{m_S^2-m_L^2}{2}+\frac{\Gamma_S^2-\Gamma_L^2}{8}\right)&
\frac{m_S^2+m_L^2}{2}+\frac{\Gamma_S^2+\Gamma_L^2}{8}\end{array}\right)
\left(\begin{array}{c}\Re_1\\ \Re_2\end{array}\right) \nonumber \\
-\left(\begin{array}{cc}\frac{\Gamma_S+\Gamma_L}{2}&
\frac{p}{q}\cdot\frac{\Gamma_S-\Gamma_L}{2}\\
\frac{q}{p}\cdot\frac{\Gamma_S-\Gamma_L}{2}&\frac{\Gamma_S+\Gamma_L}{2}
\end{array}\right)\left(\begin{array}{c}\dot\Re_1\\ \dot\Re_2
\end{array}\right)\ .\label{Eq:A.4}
\eEa
In the $\rKz\rKqz$ system, the phase of $\Gamma_{12}/m_{12}$ is small
and negative. Therefore, 
$p/q-1$ is small and positive , and we have in good approximation 
$(m_S+m_L)/2=m$, $m_S-m_L=\Delta m=-2|m_{12}|$, $(\Gamma_S+\Gamma_L)/2=\Gamma$
and $\Gamma_S-\Gamma_L=\Delta \Gamma=2|\Gamma_{12}|$. In addition,
$|\Delta m|$ and $\Gamma$ are much smaller than $m$. Neglecting
all terms that are small in second order, equation \ref{Eq:A.4} becomes
\Begeq
\left(\begin{array}{c}\ddot\Re_1\\ \ddot\Re_2\end{array}\right)=-\left(
\begin{array}{cc}m^2&m\cdot\Delta m\\m\cdot\Delta m&m\end{array}\right)
\left(\begin{array}{c}\Re_1\\ \Re_2\end{array}\right)-\left(
\begin{array}{cc}\Gamma&\frac{p}{q}\cdot\frac{\Delta\Gamma}{2}\\
\frac{q}{p}\cdot\frac{\Delta\Gamma}{2}&
\Gamma\end{array}\right)\left(\begin{array}{c}\dot\Re_1\\ \dot\Re_2
\end{array}\right)\ .\label{Eq:A.5}
\Endeq
With the substitution
\Begeq
\frac{p}{q}=1-\frac{4\Omega}{\Delta\Gamma}\ ,\label{Eq:A.6}
\Endeq
and $|\Omega| \ll |\Delta \Gamma|$
this is the equation of motion for our Foucault pendulum
with two different damping rates and two different oscillation frequencies:
\Begeq
\left(\begin{array}{c}\ddot x\\ \ddot y\end{array}\right)=-\left(
\begin{array}{cc}\omega_0^2&\omega_0\cdot \Delta\omega\\ 
\omega_0\cdot \Delta\omega&\omega_0^2 \end{array}\right)
\left(\begin{array}{c}x\\ y\end{array}\right)-\left(
\begin{array}{cc}\Gamma&\Delta\Gamma/2-2\Omega\\ \Delta\Gamma/2+2\Omega&
\Gamma\end{array}\right)\left(\begin{array}{c}\dot x\\ \dot y
\end{array}\right)\ .\label{Eq:A.7}
\Endeq

\noindent
Since the necessary approximations $|\Delta m|\ll m$, $\Gamma\ll m$ and 
$|p/q - 1| \ll 1$ are well fulfilled, our pendulum is a good representation of 
the $\rKz\rKqz$ system. The time-averaged amplitude-squares of the real parts 
in equation \ref{Eq:A.5} are equal to the modulus-squares in equation
\ref{Eq:A.1},
\Begeq
\langle 2 x^2\rangle_T =|\psi_1|^2~,~~\langle 2 y^2\rangle_T =|\psi_2|^2~,
\label{Eq:A.8}
\Endeq
with averaging-intervals $T$ which are much larger than $1/\omega_0$ 
and much smaller  than $1/\Gamma$ and $|1/\Delta \omega|$. 
Therefore, the squares of the amplitudes $x$ and 
$y$ of the pendulum represent the $\rKz$ and $\rKqz$ intensities,
respectively, in the 
evolution of $\Psi$ \cite{2012-SchubertStiewe}.\\

\noindent                                                        
{\bf V. Controls, Demonstrations, Extensions}\\

\noindent
First, we controlled the bob orbits
without rotating the table and with only one oscillation frequency by blocking
the swing hinge. Starting the pendulum in the (x+y) direction leads to
a quickly damped linear oscillation as shown in video 1 and
starting it in the direction (x-y) to the much less damped linear oscillation
in video 2. We measured the decay times 2/$\Gamma_+$ = 11.5 s
and 2/$\Gamma_-$ = 86 s, defined by the time in which the amplitudes decrease to
a level of 1/e.\\ 

\noindent
Second, we add the feature with two different pendulum lengths. This leaves 
the two motions in videos 1 and 2 essentially unchanged, but  
the oscillaion frequency in (x-y) is slightly higher than in (x+y)
because of $\Delta\omega/\Delta\Gamma < 0$. With this setup, i.\ e.\ still
with $\Omega = 0$, we next controlled the bob orbits when starting with
$(x, y, \dot x, \dot y) = (R_0,0,0,0)$ and with $(x, y, \dot x, \dot y) = 
(0,R_0,0,0)$. These motions are shown in videos 3 and 4, respectively.
Very soon after the start in video 3, we see a counterclockwise rotating 
elliptical motion with the
ellipse axes rotating clockwise. In video 4, the motion starts with 
clockwise rotating ellipses,
and the axes rotate counterclockwise. In both motions, the ellipses
become first more circle-like and then slimmer again. After about 10 s, 
the originally
counterclockwise motion changes into a slim clockwise ellipse, and vice 
versa in video 4. After about 60 s, both motions become 
linear in the (x-y)
direction, i.\ e.\ with equal coordinates x and y. The motions show energy 
transitions from the x direction to the y direction in video 3 and from
y to x in video 4. Both agree with the calculated orbits in 
figure 4,
and both are the translation of transitions from ${\rKz}$
to ${\rKqz}$
and from ${\rKqz}$ to ${\rKz}$, respectively, in the CP- and T-conserving
approximation. CP symmetry is seen in the equality of x and y at large
times.\\ 

\noindent
The complete setup with a clockwise table rotation 
($\Omega = -0,01/\rm s$) is
then demonstrated in videos 5 and 6, starting with the x motion and with the y
motion, respectively. Both motions agree with the calculated results in 
figure 5.
For early times, they resemble very much the T-symmetric
motions with $\Omega = 0$. But already at t around 10 s, we start to see
different motion patterns. The change of rotation sense is more pronounced 
in video 6 than in video 5. At later 
times, both motions end to be 
linear oscillations in the expected direction 
which is different from x = y. The x component is larger than the y component
in both motions, as given by the ratio of $\Omega/\Delta \Gamma$ in equation 7.
This is 
the translation of CP and T violation in transitions 
between ${\rKz}$ and ${\rKqz}$.
Also the message of figure 1 (c) that an initial 
${\rKqz}$ ``lives longer'' than
an initial ${\rKz}$ is clearly visible when comparing the amplitudes of the
oscillations in the two motions at the same time t greater than 50 s.\\

\noindent
The pair of coupled pendula \cite{2012-SchubertStiewe} is able 
to demonstrate CP 
violation in ${\rKz}{\rKqz}$ transitions, but in the translation of a 
quantum-mechanical
evolution which is T-symmetric and CPT-violating. The Foucault
pendulum on a rotating table, however, demonstrates CP violation in these transitions
together with T violation and CPT symmetry. All aspects of the 
${\rKz}{\rKqz}$-transition
properties in figure 1 are properly demonstrated in the Foucault-pendulum motions.\\

\noindent
With only minor modifications, we could also demonstrate 
${\rDz}{\rDqz}$, ${\rBz}{\rBqz}$ and ${\rBs}{\rBqs}$ 
transitions with our Foucault pendulum. At the present level of
experimental knowledge, CP and T violation do not play an important role in
these transitions. Since all three values of $p/q$ are compatible with 1,
we do not need a turntable. In addition, the isotropic fraction of the damping is large 
in all three systems, $\Gamma \gg |\Delta \Gamma|$, in contrast to 
the K-meson system.  
We could replace the eddy-current grid by a solid copper plate for the two 
B-meson sytems, and by a grid with nearly quadratic cells for the 
${\rDz}{\rDqz}$ system. 
The three very different $\Delta m$ values could be translated
into rigid swings at the upper end of the thread with three 
different lengths $\delta \ell$.\\

\noindent
{\bf Summary}\\

\noindent
Following a proposal of Rosner and Slezak \cite{2001-RosnerSlezak} in 2001, we have
built a Foucault pendulum on a turntable with two different oscillation 
frequencies and two different dampings. The equations of motion for the
two coordinates of this system are the real-part translations of the 
Schr\"odinger equation for the two components of the neutral K-meson
system. T violation in the K system is translated into noninvariance under
reversal-of-motion created by the Coriolis force. The consequences of
both CPT symmetry and T violation, i.\ e.\ equal probabilities for a ${\rKz}$
at $t=0$ remaining a ${\rKz}$ at any $t > 0$ and for a ${\rKqz}$ at $t=0$ 
remaining a ${\rKqz}$
at the same later time, and different probabilities for the transition
probabilities for ${\rKz} \to {\rKqz}$ and 
${\rKqz} \to {\rKz}$, become visible in the motions
of the Foucault pendulum.\\ 

\noindent
{\bf Acknowledgements}\\

\noindent
We thank Christian Enss and Karlheinz Meier for their encouragement 
and support as well as Siegfried Spiegel in the mechanical workshop of
the Kirchhoff Institute for his valuable help in assembling and testing 
our system.\\


\begin{thebibliography}{99} 
\setlength{\itemsep}{0ex}

%
\bibitem{Wu} 
         C.\ S.\ Wu\ et al., 
%       ``Experimental Test of Parity Conservation in Beta - Decay'', 
         Phys.\ Rev.\ 105 , 1413 (1957)

\bibitem{Lederman} 
         R.\ L.\ Garwin et al.,
%         ``Observations of the Failure of Conservation of Parity and Charge
%         Conjugation in Meson Decays: the Magnetic Moment of the Free Muon'',
         Phys.\ Rev.\ 105, 1415 (1957)

\bibitem{Christenson}
         J.\ H.\ Christenson et al., 
%	 ``Evidence for the 2$\pi$ Decay of the $\rKz_2$ Meson", 
	 Phys.\ Rev.\ Lett.\ 13, 138 (1964)
	 	 
\bibitem{Kobayashi}
         M.\ Kobayashi and K.\ Maskawa, Progr.\ Theor.\ Phys.\ 49, 282 (1972)	 	 
\bibitem{1970-Schubert} % T violation using unitarity 
         K.\ R.\ Schubert et al., 
%	 ``The Phase of $\eta_{00}$ and the Invariances CPT and T",\\
         Phys.\ Lett.\ 31 B, 662 (1970)
	 
\bibitem{1998-CPLEAR} % direct observation of T violation 
         A.\ Angelopoulos et al. (CPLEAR), 
%	 ``First direct observation of  time-reversal\\ non-invariance
%        in the neutral-kaon system",
         Phys.\ Lett.\ B 444, 43 (1998) 
	 
\bibitem{2006-FidecaroGerber} M.\ Fidecaro and H.-J.\ Gerber, 
%`       `The fundamental symmetries in the neutral kaon system -- a 
%        pedagogical %choice",
         Rep.\ Prog.\ Phys.\ 69, 1713  (2006)
	 
\bibitem{2010-PDG} K.\ Nakamura et al.\ (Particle Data Group),
%       ``Review of Particle Physics",\\ 
         J.\ Phys.\ G 37, 075021 (2010)
	 
\bibitem{1988-Winstein}
         B.\ Winstein,  ``CP Violation", Festi-Val -- Festschrift for Val 
         Telegdi,\\ edited
         by K.\ Winter, Elsevier Amsterdam, p.\ 245 (1988)
	 
\bibitem{1996-Rosner} 
         J.\ L.\ Rosner, 
%	 ``Tabletop time-reversal violation",
         Am.\ J.\ Phys.\ 64, 982 (1996)
	 
\bibitem{2001-KosteleckyRoberts} V.\ A.\ Kostelecky and A.\ Roberts, 
%      ``Analogue 
%         models for T and CPT violation in neutral-meson oscillations",
         Phys.\ Rev.\ D 63, 096002 (2001)
	 
\bibitem{2011-Caruso} % A gyrator circuit for demonstrating T violation
         M.\ Caruso, H.\ Fanchiotti, and C.\ A.\ Garcia Canal, 
%	 ``Equivalence
%         between classical and quantum dynamics. Neutral Kaons and electric
%         circuits", 
	 Annals of Physics 326, 2717 (2011)
	 
\bibitem{2001-RosnerSlezak} %Foucault pedulum for demonstrating T violation
         J.\ L.\ Rosner and S.\ A.\ Slezak, 
%	 ``Classical illustrations of CP 
%         violation in kaon decays",\\ 
	 Am.\ J.\ Phys.\ 69, 44 (2001)	 
	 
\bibitem{2012-SchubertStiewe} 
         K.\ R.\ Schubert and J.\ Stiewe, 
         J.\ Phys.\ G: Nucl.\ Part.\ Phys.\ 39, 033101 (2012)
%
\bibitem{BrancoLavouraSilva} 
         G.\ C.\ Branco, L.\ Lavoura, and J.\ P.\ Silva,
         ``CP Violation",\\ Oxford University Press (1999)
	 
\end{thebibliography}
\end{document}